\documentclass{article}

\usepackage{arxiv}

\usepackage[utf8]{inputenc} 
\usepackage[T1]{fontenc}    
\usepackage{hyperref}       
\usepackage{url}            
\usepackage{booktabs}       
\usepackage{amsfonts}       
\usepackage{nicefrac}       
\usepackage{microtype}      
\usepackage{cleveref}       
\usepackage{lipsum}         
\usepackage{graphicx}
\usepackage{natbib}
\usepackage{doi}

\title{DRLinFluids---An open-source python platform of coupling Deep Reinforcement Learning and OpenFOAM}

\date{}

\author{ 
    Qiulei Wang \\
	School of Civil and Environmental Engineering \\
	Harbin Institute of Technology \\
	Shenzhen 518055, China \\
	\And
	Lei Yan \\
	School of Civil and Environmental Engineering \\
	Harbin Institute of Technology \\
	Shenzhen 518055, China \\
	\And
	Gang Hu 
	\thanks{Corresponding author: hugang@hit.edu.cn} \\
	School of Civil and Environmental Engineering \\
	Harbin Institute of Technology \\
	Shenzhen 518055, China \\
	\And
	Chao Li \\
	School of Civil and Environmental Engineering \\
	Harbin Institute of Technology \\
	Shenzhen 518055, China \\
	\And
	Yiqing Xiao \\
	School of Civil and Environmental Engineering \\
	Harbin Institute of Technology \\
	Shenzhen 518055, China \\
	\And
	Hao Xiong \\
	School of Mechanical Engineering and Automation \\
	Harbin Institute of Technology \\
	Shenzhen 518055, China \\
	\And
	Jean Rabault \\
	Information Technology Department \\
	Norwegian Meteorological Institute \\
	Oslo, Norway \\
	\And
	Bernd R.\ Noack \\
	School of Mechanical Engineering and Automation \\
	Harbin Institute of Technology \\
	Shenzhen 518055, China \\
}

\renewcommand{\headeright}{}
\renewcommand{\undertitle}{}

\hypersetup{
pdftitle={DRLinFluids---An open-source python platform of coupling Deep Reinforcement Learning and OpenFOAM},
pdfsubject={cs.AI, physics.soc-ph},
pdfauthor={Qiulei Wang, Lei Yan, Gang Hu, Chao Li, Yiqing Xiao, Hao Xiong, Jean Rabault, Bernd R.\ Noack},
pdfkeywords={Deep Reinforcement Learning, OpenFOAM, Flow control, Computational fluid mechanics, and DRLinFluids},
}

\begin{document}
\maketitle

\begin{abstract}
We propose an open-source python platform for applications of Deep Reinforcement Learning (DRL) in fluid mechanics. DRL has been widely used in optimizing decision-making in nonlinear and high-dimensional problems. Here, an agent maximizes a cumulative reward with learning a feedback policy by acting in an environment. In control theory terms, the cumulative reward would correspond to the cost function, the agent to the actuator, the environment to the measured signals and the learned policy to the feedback law. Thus, DRL assumes an interactive environment or, equivalently, control plant. The setup of a numerical simulation plant with DRL is challenging and time-consuming. In this work, a novel python platform, named DRLinFluids is developed for this purpose, with DRL for flow control and optimization problems in fluid mechanics. The simulations employ OpenFOAM as popular, flexible Navier-Stokes solver in industry and academia, and Tensorforce or Tianshou as widely used versatile DRL packages. The reliability and efficiency of DRLinFluids are demonstrated for two wake stabilization benchmark problems. DRLinFluids significantly reduces the application effort of DRL in fluid mechanics and is expected to greatly accelerates academic and industrial applications.
\end{abstract}


\section{Introduction} \label{sec:intro}
\begin{figure}[h]
\centering
\includegraphics[width=0.4\textwidth]{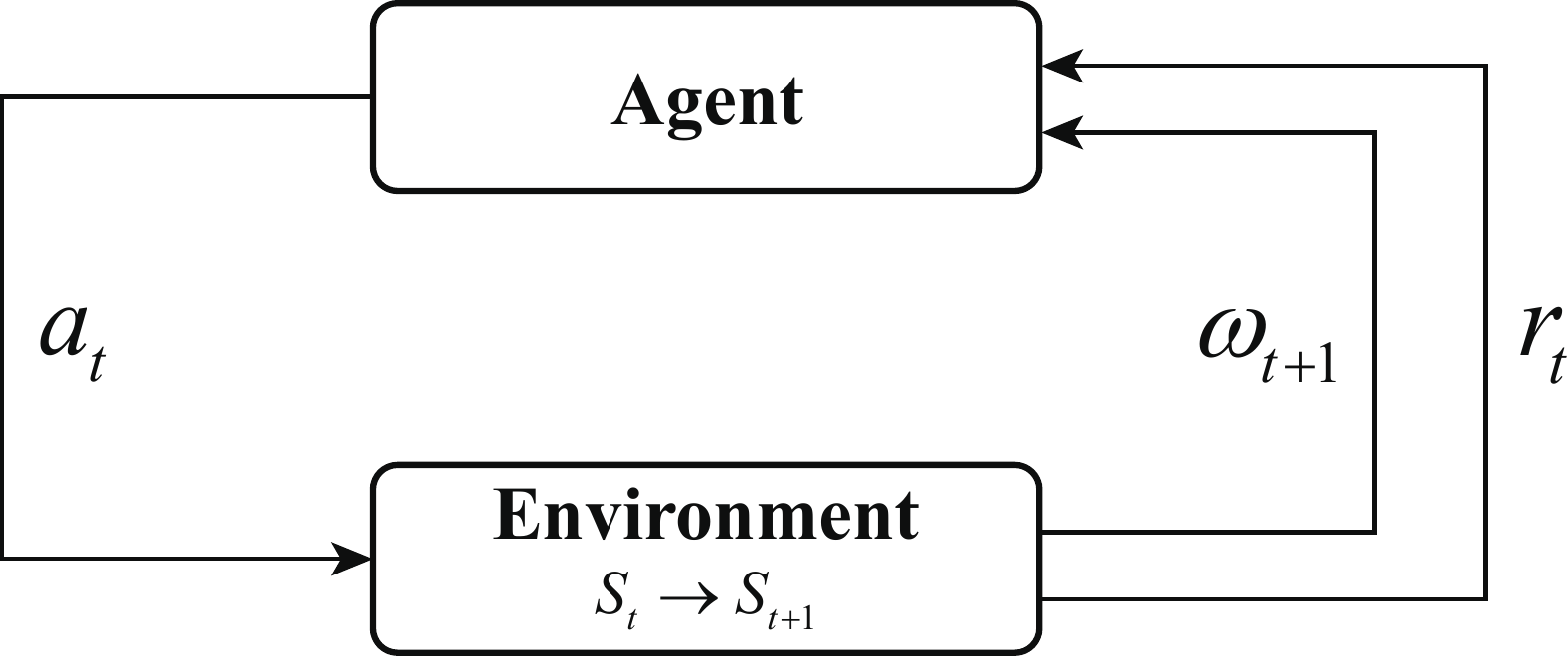}
\caption{Interaction of agent and environment in reinforcement learning} 
\label{fig:interaction}
\end{figure}

In recent years, reinforcement learning (RL), one of the machine learning paradigms, has been developed rapidly with the fast development of computer hardware and corresponding theories. Reinforcement learning is derived initially from behavioral psychology, where organisms frequently implement strategies to seek advantages and avoid disadvantages. The term \textit{reinforcement} was first introduced in 1927 by a Russian physiologist Ivan Pavlov to describe the phenomenon that specific stimuli motivate organisms more inclined to adopt certain particular strategies \citep{Pavlov1927Conditioned}. The stimulus that reinforces behavior is called reinforcer, and the strategy change resulting from reinforcer is called reinforcement learning. Inspired by behavioral psychology, Sutton proposed a formal framework for reinforcement learning \citep{Sutton1984Temporal, Sutton1988Learning}. The basic idea of RL is that the agent observes the environment, makes decisions, and gets rewards, as shown in Figure \ref{fig:interaction}. The presence of deep learning significantly accelerates the broad applications of reinforcement learning in various fields \citep{lecun2015deep}, which has been integrated very effectively into reinforcement learning with the development of deep learning \citep{mnih2015human}. The value function, policy, and model (state transition function and reward function) were approximated by a deep neural network \citep{Li2018Deep}, which forms deep reinforcement learning (DRL). A number of novel DRL algorithms were sequentially proposed, such as model-free RL methods, including A2C/A3C \citep{Mnih2016Asynchronous}, PPO \citep{Schulman2017Proximal}, DDPG \citep{Lillicrap2019Continuous}, SAC \citep{haarnoja2018soft}, DQN \citep{Mnih2013Playing}, TD3 \citep{Fujimoto2018Addressing} and model-based RL methods including World Models \citep{Ha2018World}, I2A \citep{Weber2018Imagination-Augmented}, MBMF \citep{Nagabandi2018Neural}, MBVE \citep{Feinberg2018Model-Based}, AlphaZero \citep{Silver2017Mastering} and so on.

To date, DRL has been applied widely in the field of fluid mechanics and structure engineering, such as shape optimization, active flow control, and structural reliability assessment and maintenance. \citet{Novati2017Synchronisation} investigated synchronized swimming between two self-propelled swimmers. A deep reinforcement learning algorithm was used to control swimming posture to reduce energy consumption by exploiting the vortex structure and finally improve swimming efficiency. Based on the above study, \citet{Verma2018Efficient} highlighted the influence of the past environmental state on the future state, i.e., not following first-order Markov property, using recurrent neural network to consider this effect and accelerate the learning process. The results show that energy savings could be achieved by properly utilizing the wake generated by the forward swimmer. \citet{Ma2018Fluid} proposed a DRL-based 2D coupled control system for fluids and rigid bodies, allowing a controller to drive nozzles to move on the boundary of the computational domain and to jet flow into the rigid body to accomplish a series of tasks, such as keeping the rigid body balanced, playing a two-player ping-pong game, and driving the rigid body to hit specified points on a wall in sequence. \citet{Lee2018Flow} used the DoubleDQN method to identify a series of pillars to solve the inertial flow sculpting problem, of which the DRL model has a success rate of 90\% in 200,000 episodes. \citet{xiangDeepReinforcementLearningbased2020} applied deep reinforcement learning to structural reliability assessment. using the sampling space and existing samples as states, the agent gives a set of actions along the limit state surface as the next experimental points. The results showed the advantages of the DRL-based approach with highly nonlinear problems. \citet{weiOptimalPolicyStructure2020} proposed a DRL-based framework for obtaining optimal structural maintenance strategies. The framework outputs maintenance strategies by observing bridge structures and components.

Active flow control combined with DRL may use a synthetic jet (momentum injection) or moving surface as actuation. \citet{Rabault2019Artificial} introduced DRL to active flow control for the first time by using a DRL agent model to control a pair of jets located on the upper and lower sides of a two-dimensional cylinder at low Reynolds number (\emph{Re}=100). The interactive environment was built by open-source finite-element framework FEniCS \citep{logg2012automated}, applying the PPO algorithm to learn the jet actuation strategy and  achieving significant drag reduction. Considering that the solution speed of the CFD interactive environment dramatically limits the overall progress under moderate Reynolds number conditions, a further study by \citet{Rabault2019Accelerating} performed a simultaneous training method with multiple simulation environments to shorten the training time. \citet{Ren2021Applying} applied the lattice-Boltzmann method (LBM) to establish a CFD environment with weak turbulence conditions by raising the Reynolds number to $1000$. Similar to \citet{Rabault2019Artificial}'s study, the jet actuators were deployed on the upper and lower sides of the cylinder. The results show that the DRL agent could find an effective feedback law and achieve a drag reduction of more than 30\%. On the other hand, \citet{paris2021robust} introduced the S-PPO-CMA method, a novel DRL algorithm, to optimize the sensor layout and investigated the efficiency and robustness of the identified control strategy. The proposed algorithm optimizes the sensor layout and reduces the number of sensors while keeping state-of-the-art performance. In another study, \citet{Tang2020Robust} used four synthetic jets symmetrically located on the upper and lower sides of a cylinder for active flow control. \citet{Xu2020Active} set up two small rotating cylinders behind the main cylinder obliquely with a Reynolds number of 240, and the rotational speed is controlled by a DRL agent. \citet{Fan2020Reinforcement} experimentally verified the effectiveness and feasibility of wake stabilization by two DRL-controlled rotating control cylinders. In addition, DRL has been applied to a number of chaotic active flow control tasks, including the one-dimensional Kuramoto-Sivashinski equation \citep{bucci2019control}, the one-dimensional falling fluid flow \citep{belus2019exploiting}, and the 2D Rayleigh-Benard convection \citep{beintema2020controlling}.

The shape optimization of bluff bodies has also experienced substantial progress in past decades. \citet{Garnier2021review}, inspired by \citet{Meliga2014Sensitivity}, placed a two-dimensional square cylinder in a medium Reynolds number flow field and attached a small control cylinder at a fixed position around the cylinder. Both the traditional optimization strategies and deep reinforcement learning PPO/A3C algorithm were tested to find the optimal position of the control cylinder such that the overall drag force of the two cylinders is lower than the single one. To speed up the training process, they first trained the DRL agent at a very low Reynolds number flow (\emph{Re}=10) and then applied the transfer learning technique to transfer the trained agent model to cases with higher Reynolds numbers. \citet{Viquerat2021Direct} applied DRL for the first time to direct shape optimization, using the Bezier curves technique to generate two-dimensional models with the lift-to-drag ratio of the model's aerodynamic shape as rewards, and the DRL algorithm is able to learn and generate a wing-like optimal shape without any a prior knowledge. \citet{li2021knowledge} introduced a so-called \textit{tacit} domain knowledge, such as transfer learning and meta learning, into naive DRL algorithms to obviously accelerate the process of training, which is very significant for the time-consuming interactive tasks, such as CFD simulation.

As exemplified above, DRL has been extensively and successfully applied to a variety of fluid mechanics problems, and considerable efforts are now being devoted to coupling state-of-the-art CFD solvers and DRL control algorithms, and deploying these effectively on High Performance Computing (HPC) systems \citep{2022_vinuesa_et_al}. In the aforementioned studies, the interactive CFD environments almost relied on self-programming partial differential equation (PDE) solvers based on the Lattice Boltzmann method or Navier-Stokes equation. The employed in-house solvers require significant effort and experience to achieve the efficiency and reliability of well-developed open-source CFD solvers such as OpenFOAM, in particular for complex flow fields. In contrast, OpenFOAM has been developed for more than ten years, and its accuracy and robustness have been demonstrated by substantial research \citep{ashtonVerificationValidationOpenFOAM2019,mackValidationOpenFoamHeavy2013,robertsonValidationOpenFOAMNumerical2015,suvanjumratImplementationValidationOpenFOAM2017}. Many computational fluid mechanics courses employ OpenFOAM, nowadays. Moreover, OpenFOAM is also increasingly used in industry.

Currently, there is no general and mature platform for simplifying the application of DRL in OpenFOAM simulations. This paper proposes an open-source python platform, DRLinFluids, for coupling DRL and OpenFOAM based on reliable DRL packages, including Tensorforce \citep{Kuhnle2017Tensorforce:} and Tianshou \citep{weng2021tianshou}, and OpenFOAM \citep{Jasak2007OpenFOAM:}. DRLinFluids is a flexible and scalable platform to utilize DRL in the field of computational fluid mechanics, even in continuum mechanics. Two case studies of active flow control of bluff bodies are conducted to successfully demonstrate the feasibility and reliability of DRLinFluids. We release DRLinFluids under an open source license on Github (see Appendix), and we expect that this open-source platform will greatly simplify and accelerate the relevant research and application of DRL in fluid mechanics, bluff-body aerodynamics, and wind engineering.

\section{Methodology} \label{sec:headings}

\subsection{OpenFOAM}
\begin{figure}[htb]
    \centering
    \includegraphics[width=0.6\textwidth]{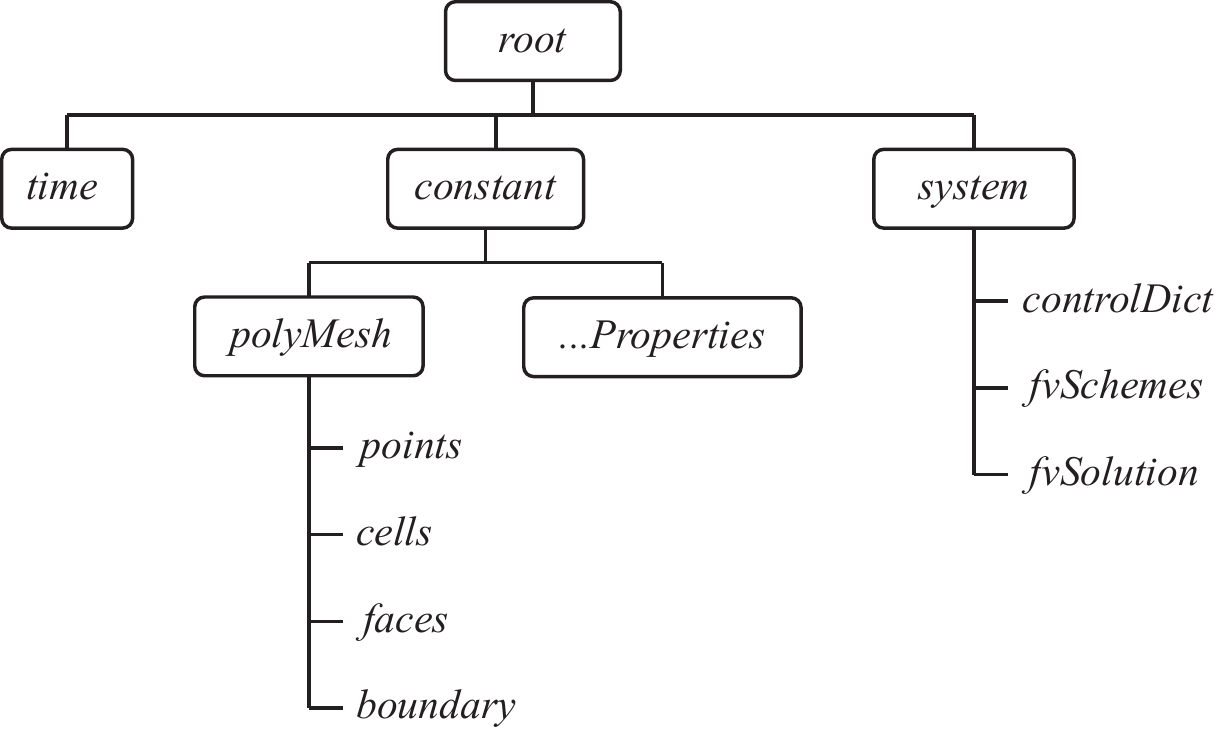}
    \caption{Directory structure of an OpenFOAM case.}
    \label{fig:directory structure}
\end{figure}

OpenFOAM is originally developed by \citet{Jasak2007OpenFOAM:}, which is an object-oriented C++ library for complex physics simulations, including structural and CFD analysis. To achieve this purpose efficiently and flexibly, OpenFOAM establishes and solves the corresponding partial differential equations (PDE) of different continuum mechanics physical phenomena with complicated models. Benefiting from the features of object-oriented programming (OOP), especially inheritance, polymorphism, and encapsulation, users can create their own solvers easily for a specific need due to the user-friendly syntax for describing partial differential equations. Besides, OpenFOAM has the following capabilities \citep{jasakOpenFOAMOpenSource2009, chenOpenFOAMComputationalFluid2014a}: (1) the ability to handle unstructured polyhedral meshes; (2) widely and reliable out-of-the-box solvers for different fields, such as computational and conjugate heat transfer, combustion, chemical reactions, multiphase flows, mass transfer, particle methods, lagrangian particles tracking, financial analysis and so on; (3) open-source with no payment and a liberal license. Figure \ref{fig:directory structure} illustrates a typically basic directory structure for an OpenFOAM case, which comprises the bare minimum of files required by running a simulation.

The aforementioned features lead OpenFOAM to an ideal solution for research and development, which is well suited as an interactive environment for reinforcement learning in complex CFD problems. 

\subsection{Reinforcement Learning}

Modern reinforcement learning algorithms are mainly based on the Markov Decision Process (MDP), which can be roughly divided into two categories, including model-based and model-free. Model-based methods can predict future states and rewards through environmental models, thereby helping agents to plan better. However, the environment model assumptions may be too idealized to reflect the internal mechanism. On the other hand, dynamic models of environments in the real world are always complex, e.g.\ the dimension of the state space is too high, and cannot be represented explicitly, making models often unavailable \cite{Brunton2015amr}. In contrast, the model-free method does not require the construction of an environment model. The agent interacts directly with the environment and improves its policy performance based on the samples obtained from the exploration. Since the model-free method neither cares about the environment model nor needs to learn the model itself, there is no problem of inaccurate environment fitting, which means that it is relatively easier to implement and train. Model-free methods can be further divided into value-based, policy-based, and combined algorithms.

The value-based methods optimize the action-value function $Q^\pi ( s,a )$. The optimal strategy $\pi^*$ can be obtained by selecting the action corresponding to the maximum value function

\begin{equation}
    \pi^*=\mathrm{arg}\mathrm{max}_\pi Q^\pi
\end{equation}

The advantage of the value-based method is that the sampling efficiency is relatively high, the estimated variance of the value function is small, and it is not easy to fall into a local optimum; the disadvantage is that it usually cannot deal with continuous action space problems, and the final strategy is usually a deterministic strategy rather than a probability distribution. In addition, the $\epsilon$-greedy strategy and the max operator in algorithms such as deep Q-networks are prone to overestimation problems. 

The policy-based method directly optimizes the policy, and maximizes the cumulative reward by iteratively updating the policy. Compared with value-based methods, policy-based methods have the advantages of simple policy parameterization, fast convergence, and are suitable for continuous or high-dimensional action spaces. As the foundations and main aspects of common DRL algorithms have been discussed extensively in both the DRL-centric literature \citep{8103164, Schulman2017Proximal, Lillicrap2019Continuous, haarnoja2018soft}, and the fluid mechanics literature \citep{Garnier2021review, rabault2020deep}, we refer the reader curious of more details about the DRL algorithms in itself to the corresponding references.

\subsection{OpenFOAM and Reinforcement Learning coupling schemes}

At present, there are many well-known deep reinforcement learning frameworks, such as DeeR \citep{FrancoisLavet2016DeeR}, Dopamine \citep{Castro2018Dopamine:}, ELF \citep{Tian2017ELF:}, OpenAI baselines \citep{Dhariwal2017OpenAI}, rllab \citep{Duan2016Benchmarking}, Keras-RL \citep{Plappert2016keras-rl}, Coach \citep{Caspi2017Reinforcement}, TF-Agents \citep{Guadarrama2018TF-Agents:}, Acme \citep{Hoffman2020Acme:}. Most of them are developed based on Python, with PyTorch and TensorFlow as automatic gradient solvers. Tensorforce \citep{Kuhnle2017Tensorforce:} and Tianshou \citep{weng2021tianshou} packages are used as an alternative ready-to-use backend to provide a wide range of reinforcement learning algorithms in the DRLinFluids platform. Besides, custom RL methods can also be easily integrated into the DRLinFluids platform by inheriting the built-in DRLinFluids's DRL class. 

To ensure DRLinFluids flexible and scalable, a low-level RL class and high-level training and evaluation platform are defined separately. The object-oriented programming (OOP) paradigm are adopted to design the RL class to improve code reusability, which organizes platform design around data and objects.

\begin{figure}[htb]
\centering
\includegraphics[width=0.9\textwidth]{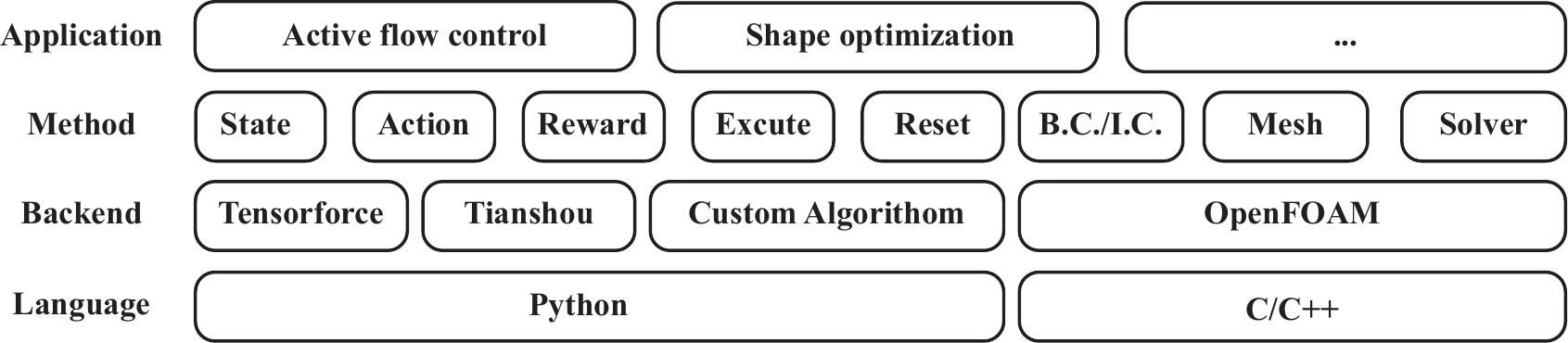}
\caption{Architecture of the DRLinFluids reinforcement learning platform.} 
\label{fig:architecture}
\end{figure}

The RL class in DRLinFluids is defined as a standard Gym environment to complete more flexible functions. Gym \citep{Brockman2016OpenAI} is a toolkit for developing and comparing reinforcement learning algorithms. It makes no assumptions about the structure of agents and is compatible with any numerical computation library, such as TensorFlow \citep{tensorflow2015-whitepaper} or PyTorch \citep{Paszke2019PyTorch:}. The architecture of DRLinFluids are shown as Figure \ref{fig:architecture}. 

\subsubsection{Low-level RL class}

The RL-OpenFOAM class takes the following parameters as input: (1) the environment parameters, including the vanilla OpenFOAM case, root path, parallel parameters, simulation dimensions, etc.; (2) state parameters, such as the type and location of probes; (3) agent parameters, such as the range of actions, interaction period between adjacent actions, and (4) learning parameters, such as actuator velocity. Member function includes execute and reset. The details of the class are shown in Table \ref{table:summary}.

\renewcommand\arraystretch{1.5}
\begin{table}
\caption{Summary of the built-in low-level RL class.} 
\label{table:summary}
\begin{tabular}{p{0.95\textwidth}}
\hline
\textbf{OpenFoam DRL Class} \\ 
\hline
\textbf{Attributes} \\
- Sensor location and identifier \\
- Controller location and identifier \\
- Controller Type and Output (action) Range \\
- Environmental parameters (e.g. inflow conditions, fluid properties, etc.) \\ 
\hline
\textbf{Method} \\
- Calling OpenFOAM operations, adding, deleting, and changing OpenFOAM configuration files \\
- DRL algorithms with Tensorforce as the backend \\ 
\hline
\textbf{Functionality} \\
- Flexible and easy to call OpenFOAM as a reinforcement learning interaction environment \\
- Suitable for a wide range of DRL Agent output (actions) types, such as jet velocity, rotating column speed, etc. \\
- Wide range of application scenarios, combined with OpenFOAM programming, to implement DRL on any PDE problem (e.g. CFD, finance, etc.) \\
- Reinforcement learning algorithms provided by Tensorforce, Tianshou, or custom algorithm. \\ 
\hline
\end{tabular}
\end{table}

\subsubsection{High-level Platform}

\begin{figure}[htb]
\centering
\includegraphics[width=0.95\textwidth]{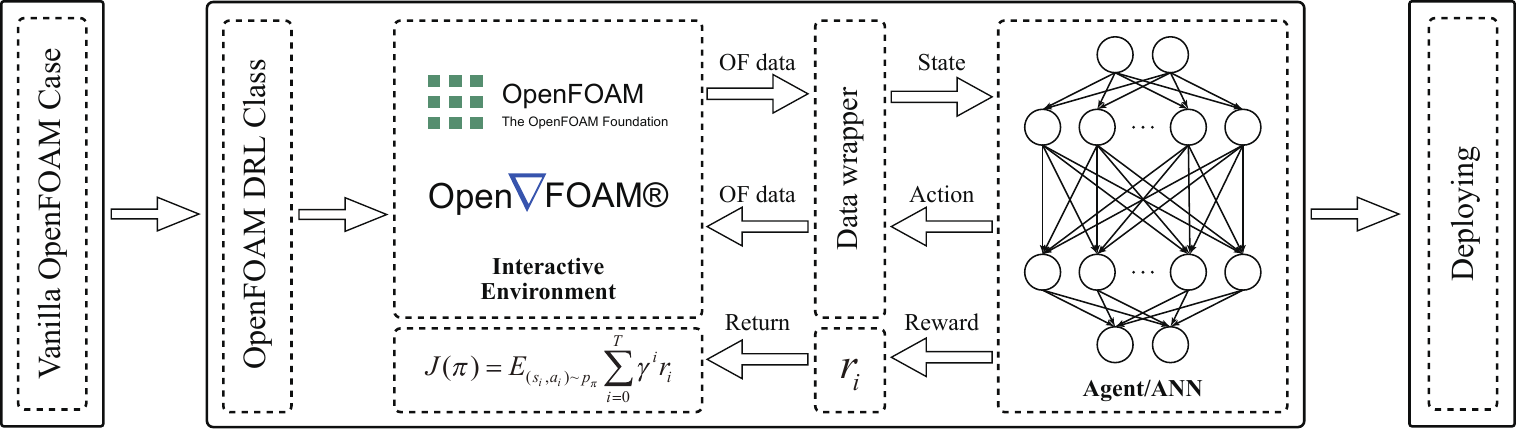}
\caption{Procedure of a reinforcement learning instance transferred from a vanilla OpenFOAM case in DRLinFluids.} 
\label{fig:procedure}
\end{figure}

To further simplify the application of DRL coupled with OpenFOAM based on the RL-OpenFOAM class, this section introduces Tensorforce and Tianshou as the backend algorithm, and builds a high-level reinforcement learning training and evaluation platform. Generally, the procedure of the reinforcement learning shown in Figure \ref{fig:procedure} can be described as the following steps:
\begin{enumerate}
    \item pre-defining a vanilla OpenFOAM simulation, including initial/boundary conditions, mesh, discrete schemes (fvSchemes), solver settings (fvSolution), and organizing these files according to the directory hierarchy specified by OpenFOAM.
    \item importing and inheriting the RL-OpenFOAM class from DRLinFluids, overwriting the reward function.
    \item setting relevant RL parameters and passing them to the RL-OpenFOAM object or instance.
    \item training, evaluating, saving, and deploying the model in practice.
\end{enumerate}

To avoid restriction introduced by OpenFOAM, this platform does not use built-in utilities. Instead, the RegExp (regular expression) and template engine are adopted to interact between OpenFOAM and Python backend according to the file input/output (I/O) stream. RegExp is a widely used method in text processing, which specifies a search pattern and uses it to search or substitute matching content. Generally, A number of ways exist to exchange or share data, such as serialization in shared memory or file I/O stream. The latter is a simple but effective method for exchanging data, writing data from memory to hard disk, and reading from hard disk to memory. It is not advisable to use this method in the HPC-like tasks because the file I/O is likely a speed bottleneck in the whole system. According to a previous CFD-based RL study \citep{Rabault2019Accelerating} and our experience which will be further discussed in section \ref{sec:case studies}, the CFD simulation part accounts for over 99\% of the total computation time. Therefore, the I/O stream is highly suitable for CFD-based RL tasks, in which CFD simulation is very time-consuming compared to the communication in the I/O stream, even agent policy optimization.

\begin{figure}[htb]
\centering
\includegraphics[width=0.95\textwidth]{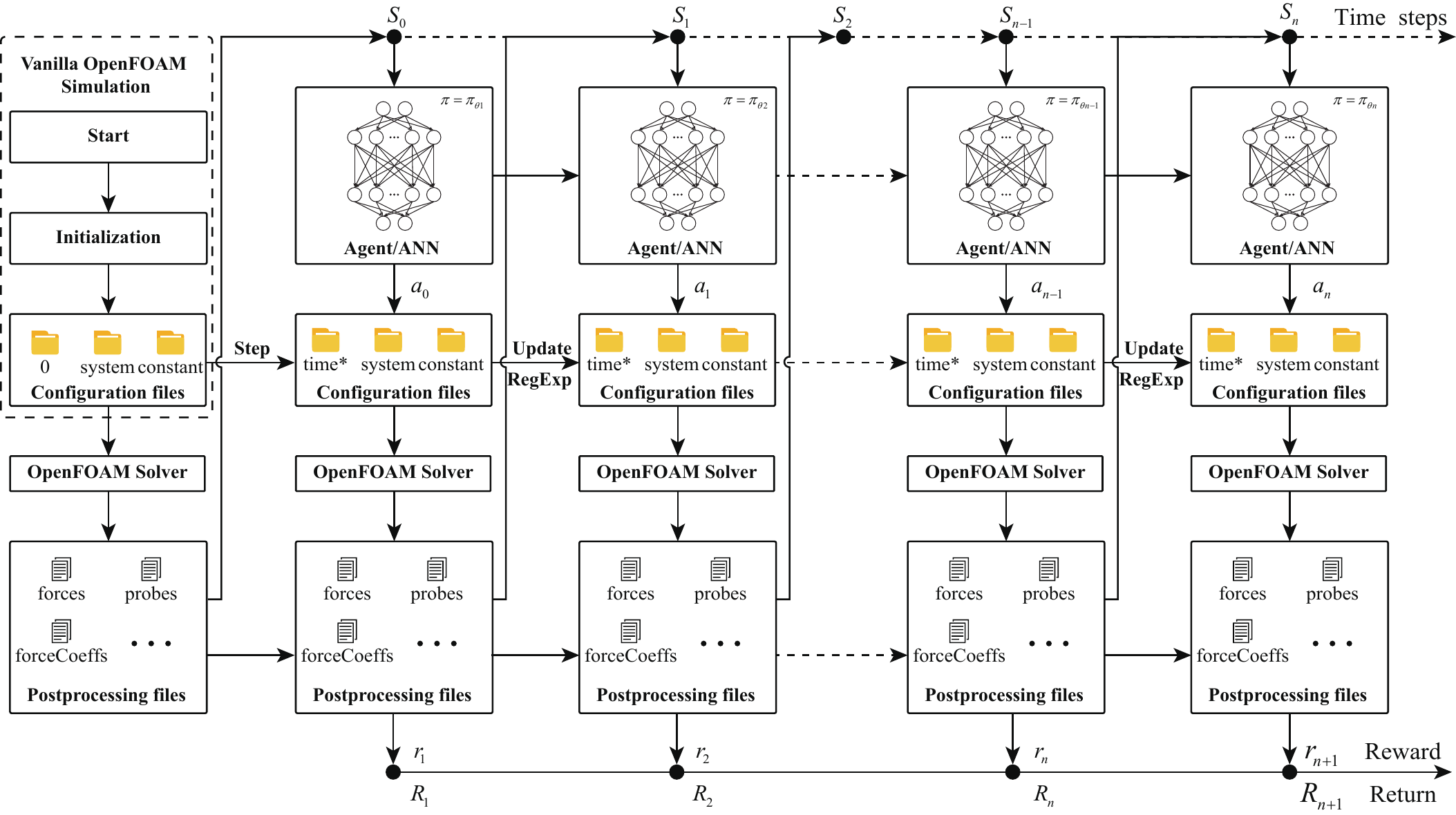}
\caption{Detail procedures inside DRLinFluids.} 
\label{fig:steps}
\end{figure}

The interaction details inside DRLinFluids are shown in Figure \ref{fig:steps}, which is also a standard procedure from scratch. The DRLinFluids will call the OpenFOAM solver to generate an initial environment with states at the beginning. Then the loop of interaction between agent and environment, and policy network optimization are carried out step by step.

\section{Case studies} \label{sec:case studies}
To demonstrate the feasibility and reliability of DRLinFluids, two case studies of flow control of bluff bodies are tested based on DRLinFluids. The first case is an adaptation of a circular cylinder flow control conducted by \citet{Rabault2019Artificial}. The second case is a square cylinder flow control by jet flow at trailing edges of two sides. The successes of DRLinFluids in these two cases well prove the feasibility and reliability of DRLinFluids. The details of these case studies are documented below. 
\subsection{Case 1: Active flow control of a 2D circular cylinder}
Flow around a circular cylinder has been studied extensively in the past century. When flow passes a circular cylinder, a series of vortexes shed from its two sides and form the well-known K\'arm\'an vortex street. The vortex shedding results in increased aerodynamic drag and unsteady lift forces acting on the cylinder, both of which are normally not desirable. In this study, two synthetic jets are located on the upper and lower side of the circular cylinder and controlled by DRL algorithm in order to minimize the drag and lift forces of the cylinder in an energy-efficient way. The whole process of this study is conducted based on DRLinFluids.

\subsubsection{Numerical Simulation Model}

\begin{figure}[htb]
\centering
\includegraphics[width=0.9\textwidth]{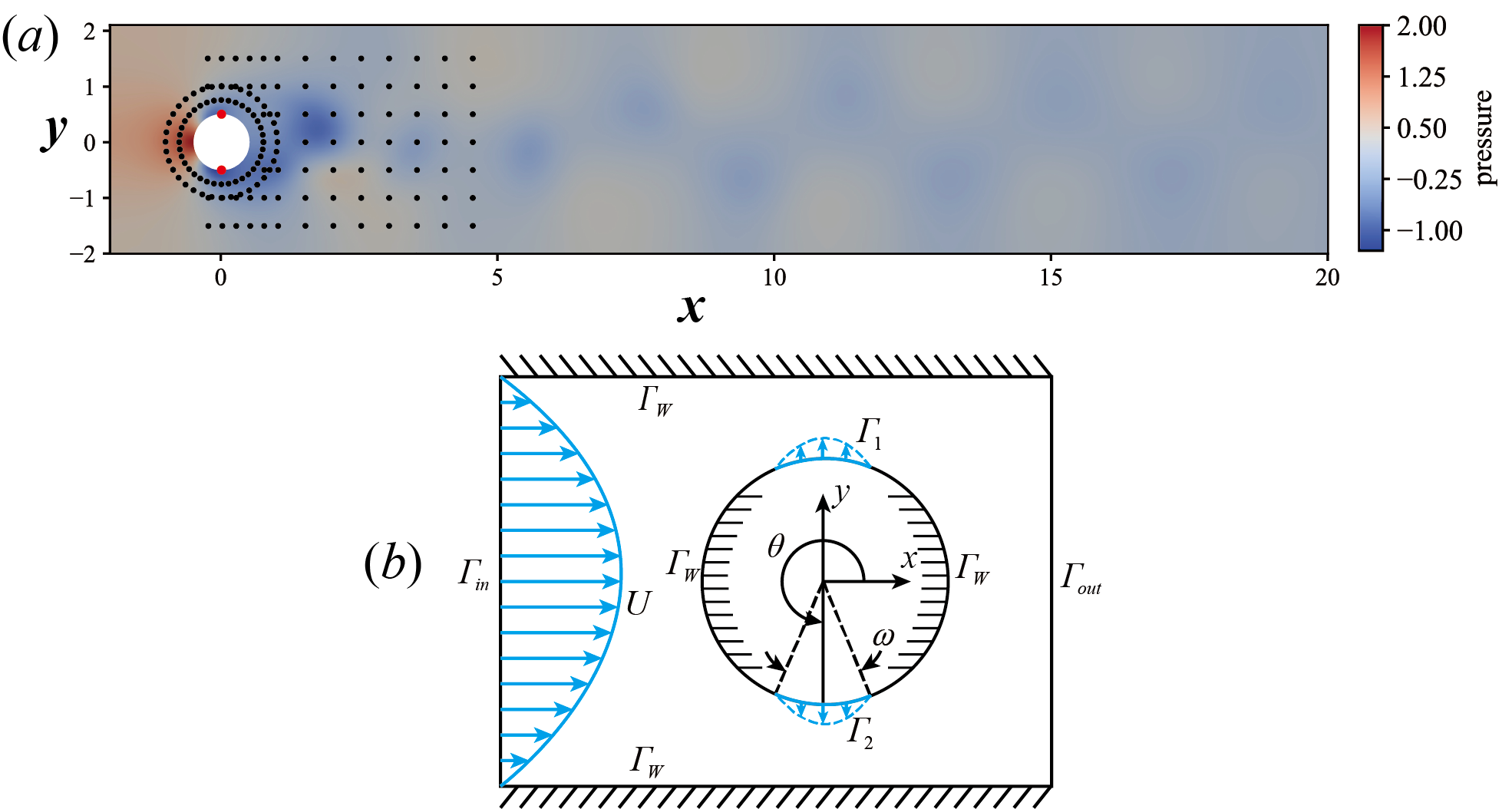} 
\caption{Description of the numerical setup adapted from \citet{schafer1996benchmark}: (a) unsteady pressure field around the circular cylinder after flow initialization without active control. The locations of the 149 velocity probes are indicated by the black dots. The locations of 2 jets are indicated by the red dots; (b) details of the cylinder and the jet actuators.} 
\label{fig:probes}
\end{figure}

The simulation environment is built based on OpenFOAM, and unstructured meshes are adopted for the CFD simulation. The mesh consists of 16200 triangular elements with increased resolution near the cylinder. The diameter of the cylinder is $D$. The length and width of the computational domain are $L = 22D$ and $B = 4.1D$ respectively as shown in Figure \ref{fig:probes}(a). Similar to the benchmark case \citep{schafer1996benchmark}, the cylinder is located slightly away from the centre of the computational domain, in order to trigger the vortex shedding. The distance from the outlet to the leeward side of the cylinder is $19.5D$, so that the wake can fully develop. Using a non-dimensional time step $dt = 5\times 10^{-4}$, the CFL number ( $\frac{\mathrm{d}t|U|}{\delta x}\ $, $|U|$ is the modulus of the velocity vector in a grid unit, ${\delta x}$ is the grid length towards the velocity direction) is less than 1, which meets the accuracy requirements.

The inlet boundary $\Gamma_{in}$ uses the velocity inlet boundary condition, and the inflow velocity is set to a parabolic channel flow profile with a mean velocity magnitude of $U$. The corresponding Reynolds number $Re$ is 100 ( $Re = \frac{UD}{\nu}\ $, $\nu$ is the kinematic viscosity). An outflow boundary condition $\Gamma_{out}$ is set for the outlet of the domain. The non-slip wall boundary condition $\Gamma_W$ is applied on the top and bottom of the channel, as well as on the cylinder surfaces (see Figure \ref{fig:probes}(b)).

The flow control action is performed by two jet holes ($\Gamma_1$ and $\Gamma_2$) at two sides of cylinder. A parabolic velocity distribution with a jet width of $\omega = 10^\circ$ is imposed at these two jet holes, as shown in Figure \ref{fig:probes}(b). The direction of the jet actuators is perpendicular to the cylinder wall, which implies that the drag reduction results from the indirect flow control, rather than direct injection of momentum. Meanwhile, the sum of the velocities of the jets is zero, i.e. $ V_{\Gamma_1} = -V_{\Gamma_2}$.

The probes sensing the state for this case can utilize either velocity or pressure probes. For this case, only velocity probes are setup and they are arranged in two circles around the cylinder and also in the wake region with a total of 149 probes, as shown in Figure \ref{fig:probes}(a). These probes monitor the horizontal and vertical components of the velocity field, and hence the agent can perceive detailed information of the flow field and learn from vortex shedding patterns.

\subsubsection{Formulation of optimizing flow control}

The feedback stabilization of a circular cylinder is a typical optimization problem and leads reduced drag and vanishing lift forces. The use of DRL algorithms is essentially to find a control law that optimizes a given reward function in order to minimize drag and lift forces. This control law can make the reward function show a certain upward trend when it is properly trained. The proposed reward function is constructed as a combination of drag and lift coefficients in this study. The reward function is set as follows:
\begin{equation}
    r_t = (C_D)_{T_0} - (C_D)_T - 0.1|(C_L)_T| 
\label{con:rewardwfunction}
\end{equation}
where $(C_D)_{T_0} = 3.205$ is the mean drag coefficient of the circular cylinder without flow control, $(\cdot)_T$ representing the sliding average over the duration of one jet flow control period $T$ corresponding to active flow control cylinder.

The reward function in Equation (\ref{con:rewardwfunction}) has been verified to be better than simply using the instantaneous drag coefficient, i.e. $r_t = - (C_D)_T$. First, only considering the change in drag coefficient could cause the lift to fluctuate too much, the algorithm tends to make the control value run to the extreme value. Furthermore, it has been proved that adding lift can improve the learning speed and stability. Second, the drag reduction is defined as $\Delta C_D = (C_D)_{T_0} - (C_D)_T$ relative to uncontrolled flow. This equation is designed to be as positive as possible, and as the training effect gets better, the reward function tends to increase while the drag value decreases. Meanwhile, it is necessary to keep changes in the drag ($\Delta C_D$) and in the lift ($\Delta C_L = 0.1|(C_L)_T|$) the same order of magnitude, so that the algorithm does not only focus on drag reduction and ignore changes in lift during training. After taking the above points into account, the reinforcement learning algorithms could try to maximize $r_t$, by reducing both drag and lift during training.

The jet actuation is smoothed to ensure a continuous changing control signal without excessive lift fluctuations due to a sudden change in the jet velocity. To this end, the control action is then adjusted from one non-dimensional time step in the simulation to the next by
\begin{equation}
    V_{\Gamma_1(t)} = V_{\Gamma_1(t-1)}+ \alpha[a - V_{\Gamma_1(t-1)}]
\end{equation}
where $\alpha = 0.1$ is a numerical parameter determined by trial and error, $V_{\Gamma_1(t)}$ and $V_{\Gamma_1(t-1)}$ is the jet flow velocity adopted at non-dimensional time instant $t$ and $t-1$ respectively, and $a$ is one jet flow velocity for the current 50 time steps proposed by the DRL agent.

\subsubsection{DRL algorithm configuration and simulation results}

The DRL framework uses the DRLinFluids framework, and the agent is trained by using the Proximity Policy Optimization (PPO) method, which is loaded from Tensorforce platform  \citep{Kuhnle2017Tensorforce:}. The PPO method is a policy gradient algorithm, and its training alternates between sampling data interactively with the environment and optimizing a surrogate objective function by using stochastic gradient ascent. It is one of the most commonly used algorithms in DRL \citep{Schulman2017Proximal}. The ANN architecture consists of two dense layers with 512 fully connected neurons, the input layer receives data from velocity probes, and the output layer produces a jet velocity. The training loop of DRL is shown in Figure \ref{fig:procedure}. Training is performed using a parallel environment, where the initial state of each environment is obtained by performing simulations without active control until a fully developed wake with a Karman vortex street. The state of the flow field at this point is stored and used as the starting point for subsequent learning episodes. At the beginning of the learning process, the PPO agent interacts and updates the ANN hyperparameters every 50 time steps. This numerical trick and the smoothing strategy described in Eq. (\ref{con:rewardwfunction}) together ensure a continuous control signal, which is crucial for a successful DRL flow control process.

\begin{figure}[htb]
\centering
\includegraphics[width=0.5\textwidth]{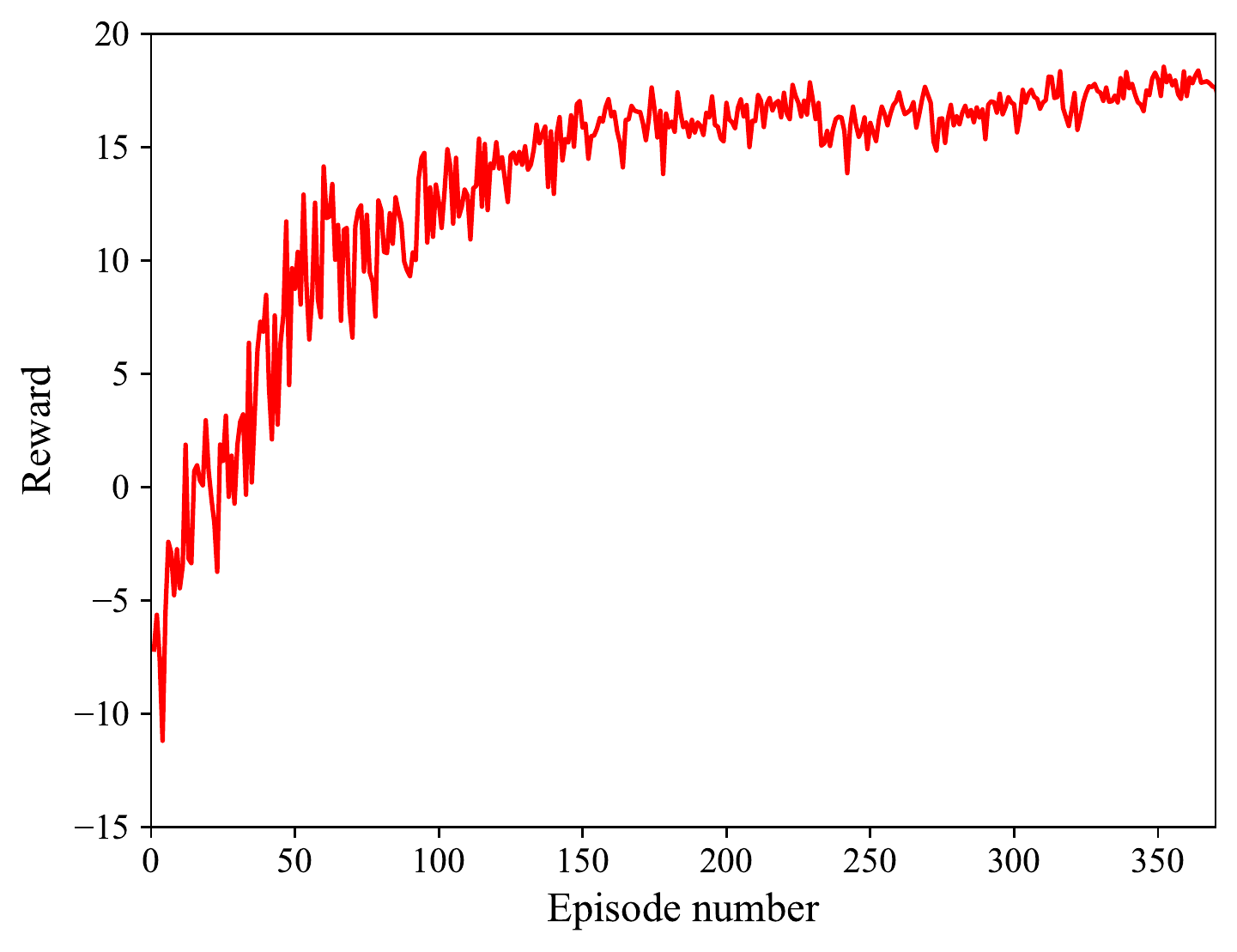} 
\caption{Reward curve from parallel environment of DRL-based active flow control for circular cylinder.}
\label{fig:cylinderreward}
\end{figure}

Using those methods, and choosing an episode duration period $T_{\rm max} = 2.5$ (corresponds to 5000 numerical non-dimensional time steps, i.e.\ 100 jet control flow set by the agent), the PPO agent is able to learn a control strategy through parallelizing the data sampling from the epochs. In Figure \ref{fig:cylinderreward}, it is a single training environment reward development process corresponding to 4 environments parallel training. After approximately 200 epochs, the reward is closed to the maxima. Fully stabilized control strategy is obtained with further epochs so that the purpose of tuning the policy would be achieved. Furthermore, it has been proved that no more improvement is obtained if the network is allowed to train for a longer duration. 

\begin{figure}[htb]
\centering
\includegraphics[width=0.9\textwidth]{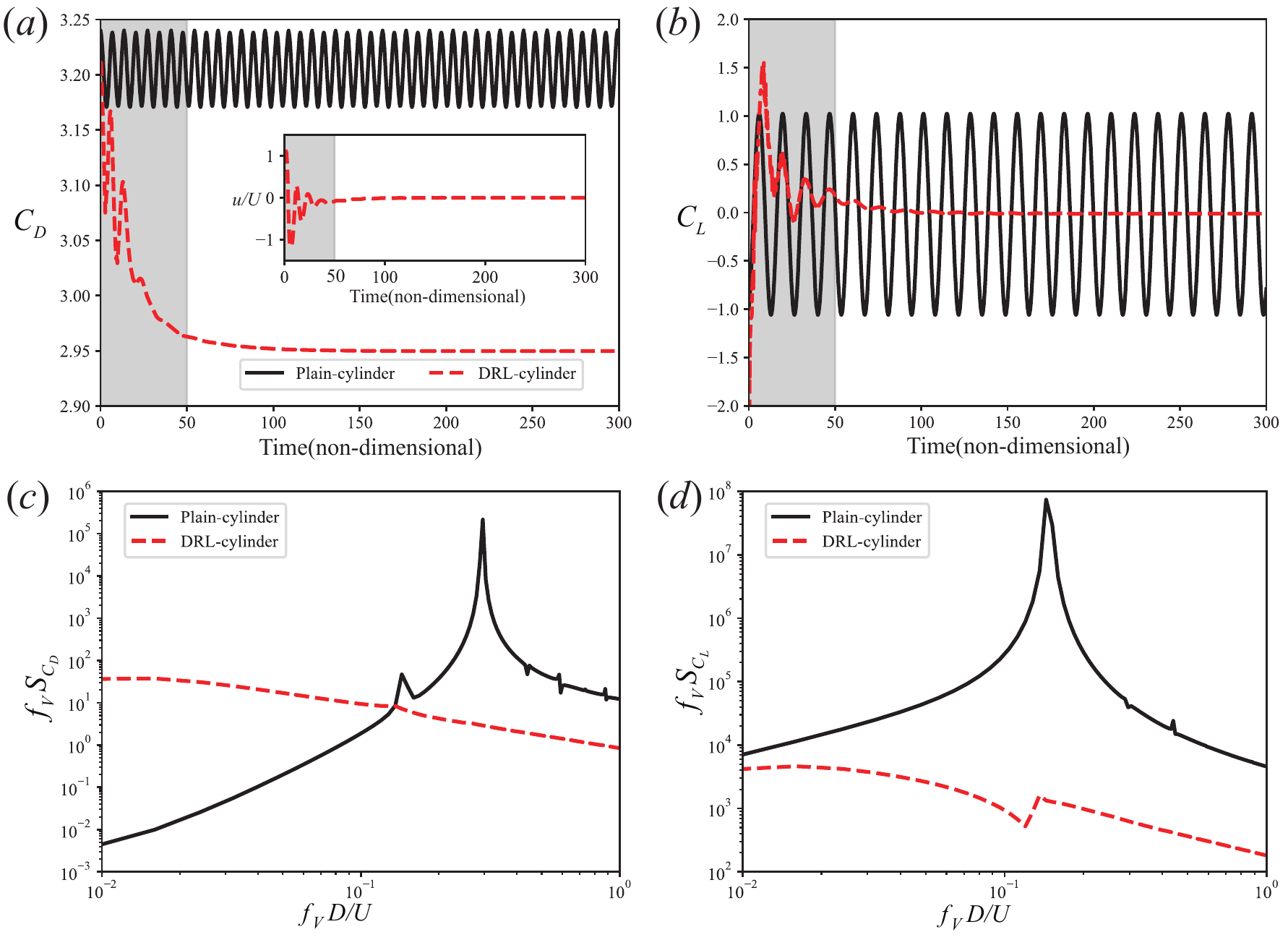} 
\caption{(a) Time-resolved value of drag coefficient $C_D$ for the cylinder without (Plain-cylinder) and with (DRL-cylinder) active flow control, and corresponding normalized velocity flow rate of the jet flow $\Gamma_1$; (b) Time-resolved value of the lift coefficient $C_L$ for the cylinder without (Plain-cylinder) and with (DRL-cylinder) active flow control; (c) power spectral density of the drag coefficient $C_D$ during the period of non-dimensional time ranging from 50 to 300; (d) power spectral density of the lift coefficient $C_L$ during the period of non-dimensional time ranging from 50 to 300.}
\label{fig:cdcl}
\end{figure}

As shown in Figure \ref{fig:cdcl}, the artificial neural network has been successfully trained by the PPO algorithm in the DRLinFluids framework and is able to perform active flow control to continuously reduce drag and suppress lift. As shown in Figure \ref{fig:cdcl}(a), in the absence of actuation, the drag coefficient $C_D$ oscillates periodically around a mean value. The mean value of drag coefficient is 3.205, and its standard deviation value is 0.0236. The standard deviation value of the lift coefficient is 0.74. With DRL-based active flow control (denoted by DRL-cylinder), the mean drag coefficient of the cylinder is reduced to 2.95, corresponding to a drag reduction of approximately 8$\%$, which is the optimal drag reduction attainable in this configuration, as discussed in \citet{Rabault2019Artificial}. In addition, the fluctuation of the drag coefficient is also significantly suppressed, corresponding to the standard deviation value of 0.0028, only 11.86\% of that the unforced cylinder. On the other hand, in Figure \ref{fig:cdcl}(b), the lift coefficient fluctuates at the very early stage, while the fluctuation is almost fully suppressed when a proper control strategy is found after a certain number of iterations of training. The standard deviation value of the lift coefficient is significantly reduced to 0.032, only 4.32$\%$ that of the plain cylinder, which implies that the vortex shedding is almost fully suppressed.

Power spectrum analyses of $C_D$ and $C_L$ of the cylinder with and without active flow control are conducted and presented in Figure \ref{fig:cdcl}(c) and (d). An obvious peak can be observed from the power spectrum curves for both $C_D$ and $C_L$ of the plain cylinder, which implies regular vortex containing considerable energy and shedding from the plain cylinder. In contrast, the peaks disappear in the power spectrum curves of $C_D$ and $C_L$ of the cylinder with active flow control. That is to say, the regular vortex shedding has been destroyed by the jet flow.   

\begin{figure}[htb]
\centering
\includegraphics[width=0.7\textwidth]{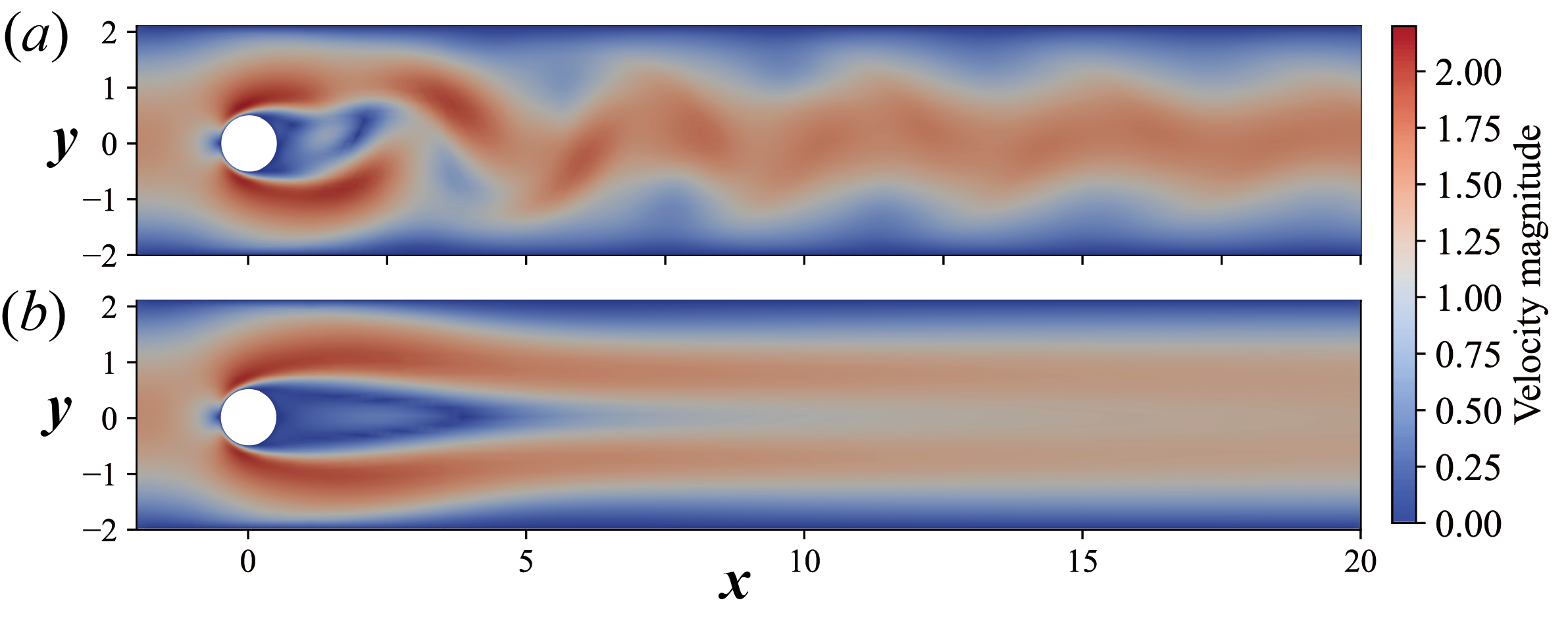} 
\caption{Comparison of representative snapshots of the velocity field around the cylinder (a) without and (b) with active flow control.} 
\label{fig:cylinder_comparisonU}
\end{figure}

Figure \ref{fig:cylinder_comparisonU} presents the instantaneous flow field of the circular cylinder with or without active control. The effect of the controlled jet flow on the mitigation of aerodynamic force of the cylinder could be further elaborated in terms of the flow pattern. In Figure \ref{fig:cylinder_comparisonU}(a), an alternative vortex shedding pattern is clear for the plain cylinder as expected. As is well known, this alternative vortex shedding pattern directly results in the fluctuations in both drag and lift coefficients which shown in Figure \ref{fig:cdcl}. In terms of the evolution of both force coefficients and the jet velocity shown in Figure \ref{fig:cdcl}, the flow control process can be divided into two stages. First, the DRL agent modifies the vortex shedding by imposing a relatively large jet flow in the period of non-dimensional time ranging from 0 to approximately 50, which reduces drag and lift considerably. Second, after this unstable control stage, the jet flow fluctuates weakly and gradually approaches zero. Only a very weak jet flow is required in the later stage when the vortex shedding is stabilized. The corresponding ratio of the jet speed to the incoming wind speed $u/U$ is only 0.58$\%$. In Figure \ref{fig:cylinder_comparisonU}(b), with small actuation, the alternative vortex shedding is suppressed, leading to the reduction in the fluctuations of both $C_D$ and $C_L$. In the mean time, an elongated re-circulation zone forms in the near wake. This elongated zone is associated with increasing pressure in the wake and hence reduces the drag force of the cylinder. Apparently, the DRL agent has learned an effective and efficient strategy for controlling the jet flow for minimizing the drag and lift of the cylinder.

\subsection{Case 2: Active flow control of a 2D square cylinder}

A cylinder with a square cross section has also been widely used in practical engineering such as tall buildings, bridge pylons. The square cylinder is also a classical bluff body suffering from strong flow-induced instability. Vortex shedding from the square cylinder induces alternative force acting on the cylinder, and this force is able to cause strong crosswind vibrations and hence potential structural failure. As a result, mitigation of aerodynamic forces of square cylinders has attracted substantial research in the past. In this study, two synthetic jets are located on the trailing edges of two sides of the square cylinder and intelligently controlled by DRL in order to minimize its drag and lift in an energy-efficient manner. The entire training process is implemented using DRLinFluids.

\subsubsection{Numerical Simulation Model}

\begin{figure}[htb]
\centering
\includegraphics[width=0.7\textwidth]{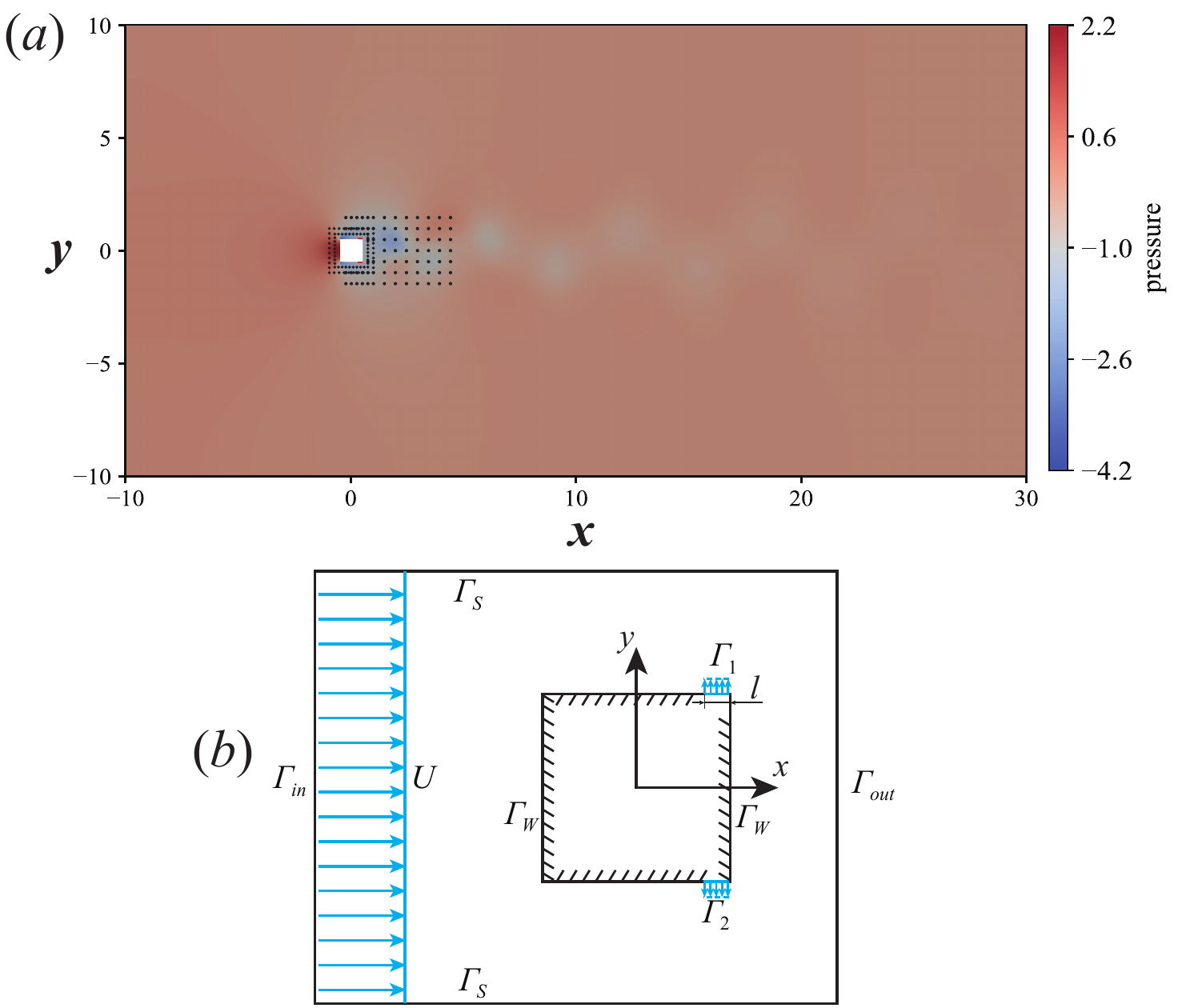} 
\caption{Description of the numerical setup: (a) unsteady pressure field around the square cylinder after flow initialization without active control. The location of the velocity probes is indicated by the black dots. The location of the control jets is indicated by the red dots; (b) details of the square cylinder and the jet actuators.} 
\label{fig:squareprobes}
\end{figure}

The length/width of the square cylinder is $D$, and the length and width of the computational domain are $L = 40 \>D$ and $B = 20 \>D$, respectively, as shown in Figure \ref{fig:squareprobes}(a). The distance from the outlet to the leeward face of the square cylinder is $29.5\> D$, thus minimizing the effect of the outflow condition on vortex shedding. Structural meshes with 23125 quadrilateral elements are adopted for the CFD simulation, and they are refined around the square cylinder. A non-dimensional time step $dt = 5\times 10^{-4}$ is adopted. The corresponding CFL number is less than unity, which meets typical accuracy requirements.

At the inlet boundary $\Gamma_{in}$ uniform flow  with velocity $U$ is imposed. The corresponding Reynolds number $Re$ is 100. An outflow boundary condition $\Gamma_{out}$ is set for the outlet of the domain. A symmetrical boundary condition $\Gamma_S$ is applied to the top and bottom of the domain, and a non-slip wall boundary condition $\Gamma_W$ is adopted for the surface of the square cylinder (see Figure \ref{fig:squareprobes}(b)).

The flow control action is implemented by two jet flows ($\Gamma_1$ and $\Gamma_2$) located on the trailing edges of two sides of the square cylinder. A uniform flow velocity distribution with a jet width $l = D/25$ is applied at these two jet holes, as shown in Figure \ref{fig:squareprobes}(b). Similar to the first case study, the sum of the jet velocities is zero, i.e. $ V_{\Gamma_1} = -V_{\Gamma_2}$. The state of the DRL model is set as flow velocities around the cylinder monitored by a total of 147 velocity probes distributed in the vicinity of the cylinder as shown in Figure \ref{fig:squareprobes}(a). 

\subsubsection{DRL algorithm configuration and simulation results}

The DRLinFluids framework is used as the DRL framework. The agent is trained using the Soft Actor-Critic (SAC) method, which is loaded from Tianshou Deep Reinforcement Learning Library\citep{weng2021tianshou}. SAC is an off-policy actor-critic DRL algorithm based on a maximum entropy reinforcement learning theory. The actor's goal is to maximize expected reward as well as maximize entropy, and succeed in reaching the desired value while acting as randomly as possible. Since it is an off-policy algorithm, training can be performed efficiently with limited samples. More details can be found in \citet{haarnoja2018soft}. The ANN architecture also consists of two dense layers of 512 fully connected neurons, identical to that in the first case study. The input layer receives data from those velocity probes, and the output layer is the jet velocity. Since the vortex shedding frequency of square cylinder is higher, the SAC agent interacts and updates the ANN hyperparameters every 26 time steps. Training is performed using 5 parallel environments, each with an initial state of a fully developed stable wake and used as a starting point for subsequent training.

\begin{figure}[htb]
\centering
\includegraphics[width=0.5\textwidth]{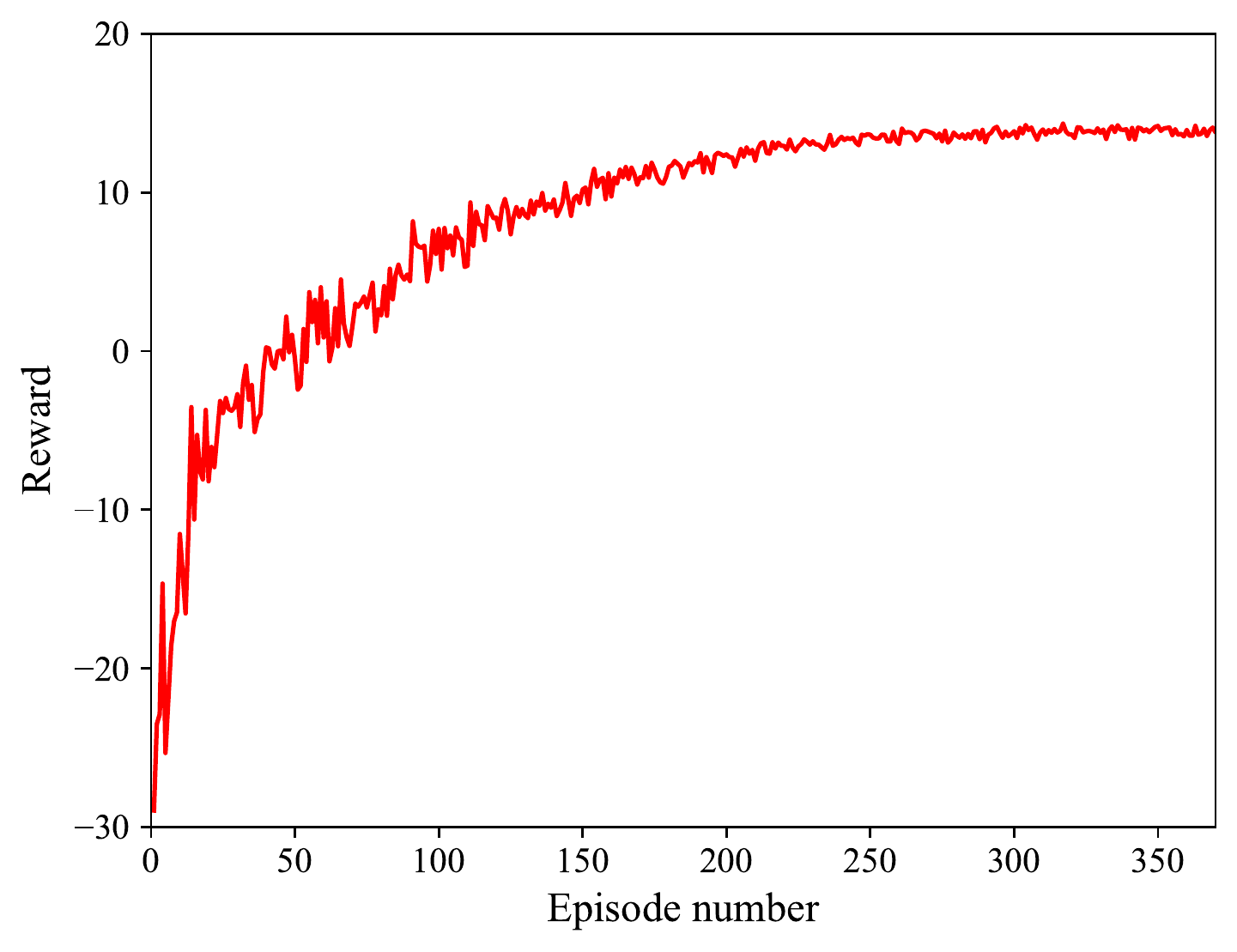} 
\caption{Reward curve from parallel environment of DRL-based active flow control for square cylinder.}
\label{fig:squarereward}
\end{figure}

Through choosing an episode duration period $T_{\rm max} = 1.3$ (corresponds to 2600 numerical non-dimensional time steps, i.e. 100 jet flow set by the agent), the SAC agent is able to learn a control strategy through parallelizing the data sampling from these epochs. In Figure \ref{fig:squarereward}, it is a single training environment reward development process corresponding to 5 environments parallel training. After typically approximately 250 epochs, the reward reaches its peak value, demonstrating that a stabilized control strategy is obtained. In fact, considerable computation time is spent in the flow simulation. Such a simple and fast simulation setup makes reproduction of the results easily.

\begin{figure}[htb]
\centering
\includegraphics[width=0.9\textwidth]{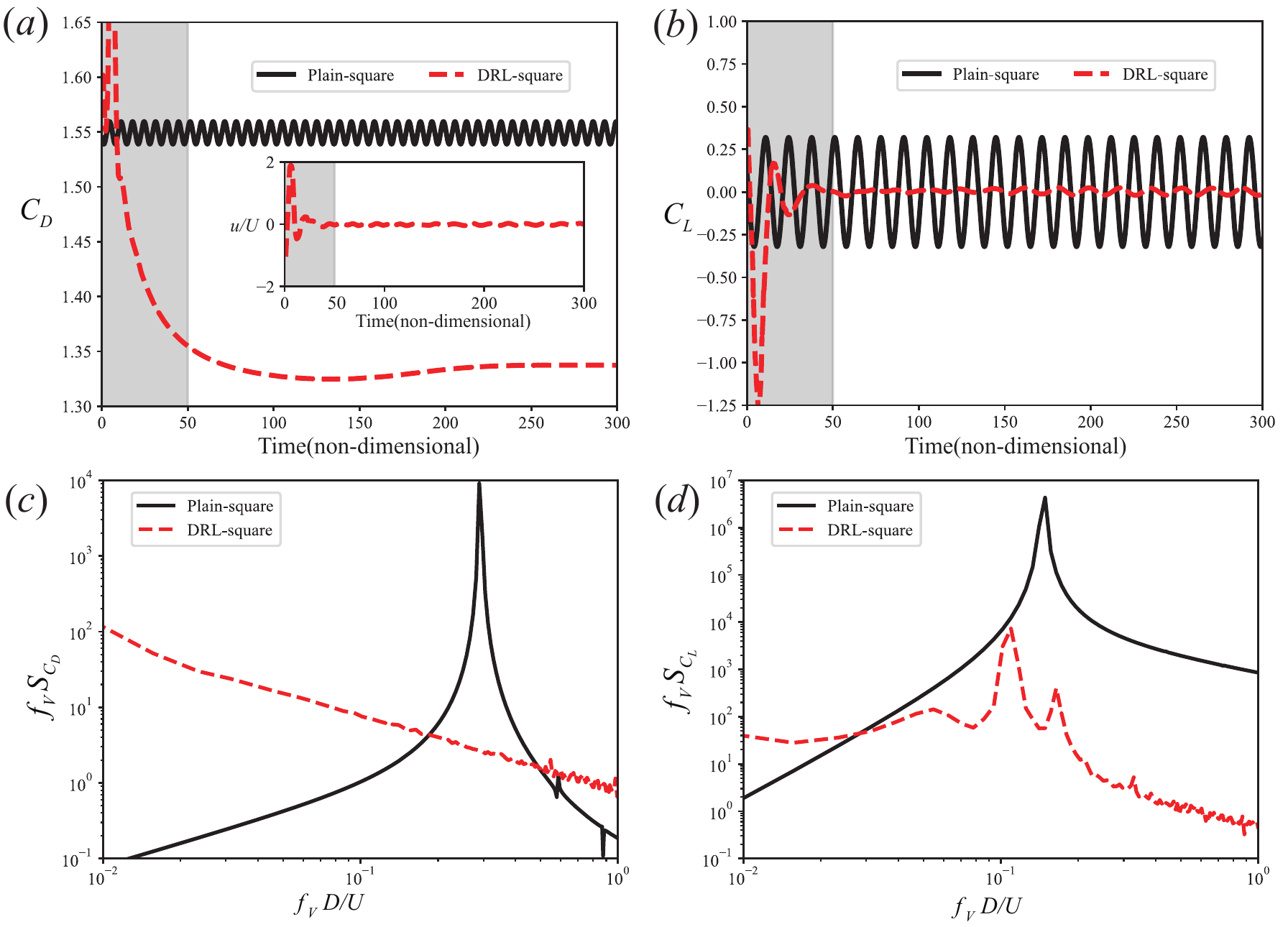} 
\caption{(a) Time-resolved value of the drag coefficient $C_D$ of the square cylinder without (Plain-square) and with (DRL-square) active flow control, and corresponding normalized velocity flow rate of the control $\Gamma_1$; (b) Time-resolved value of the lift coefficient $C_L$ of the square case without (Plain-square) and with (DRL-square) active flow control; (c) power spectral of the drag coefficient $C_D$ during the period of non-dimensional time ranging from 50 to 300. (d) power spectral of the lift coefficient $C_L$ during the period of non-dimensional time ranging from 50 to 300.} 
\label{fig:squarecdcl}
\end{figure}

As shown in Figure \ref{fig:squarecdcl}, it is clear that the artificial neural network has been successfully trained by the SAC algorithm in the DRLinFluids framework to intelligently control the jet flow in order to minimize the aerodynamic forces of the square cylinder. For the plain square cylinder, the mean drag coefficient is 1.549, and its standard deviation value is 0.0073. The standard deviation value of the lift coefficient is 0.227. With the active flow control, the mean drag coefficient of the cylinder is reduced to 1.337, corresponding to a drag reduction of approximately 13.7$\%$. In addition, the standard deviation of the drag coefficient is decreased to 0.0061. More importantly, the standard deviation of the lift coefficient is significantly reduced to 0.0134, only 5.9$\%$ of that the plain square cylinder. The suppression of the lift force is desirable for the mitigation of flow-induced instability of the square cylinder.

Power spectrum analyses of $C_D$ and $C_L$ of the square cylinder with and without active flow control are conducted and presented in Figure \ref{fig:squarecdcl}(c) and (d). An obvious peak value can be observed from the power spectrum curves for both $C_D$ and $C_L$ of the plain cylinder, which shows regular vortex containing considerable energy and shedding from the plain square. On the contrary, with active flow control, the peak is eliminated in the power spectrum curves of $C_D$ of the square cylinder. It shows that the regular vortex shedding has been fully disrupted by the jet actuation.

Two successive states can be observed from the time histories of aerodynamic forces and jet velocities shown in Figure \ref{fig:squarecdcl}(a). First, the DRL agent performs relatively large instantaneous jet to greatly reduce drag during the period of non-dimensional time ranging from 0 to approximately 50. Afterward, the controlled jet flow is stable with a very small velocity. The corresponding ratio of the peak jet velocity to the incoming wind speed $u/U$ is only 3.443$\%$. Apparently, the DRL agent has found an effective and efficient solution to maintain a small drag and a very weak lift for the square cylinder.

\begin{figure}[htb]
\centering
\includegraphics[width=0.7\textwidth]{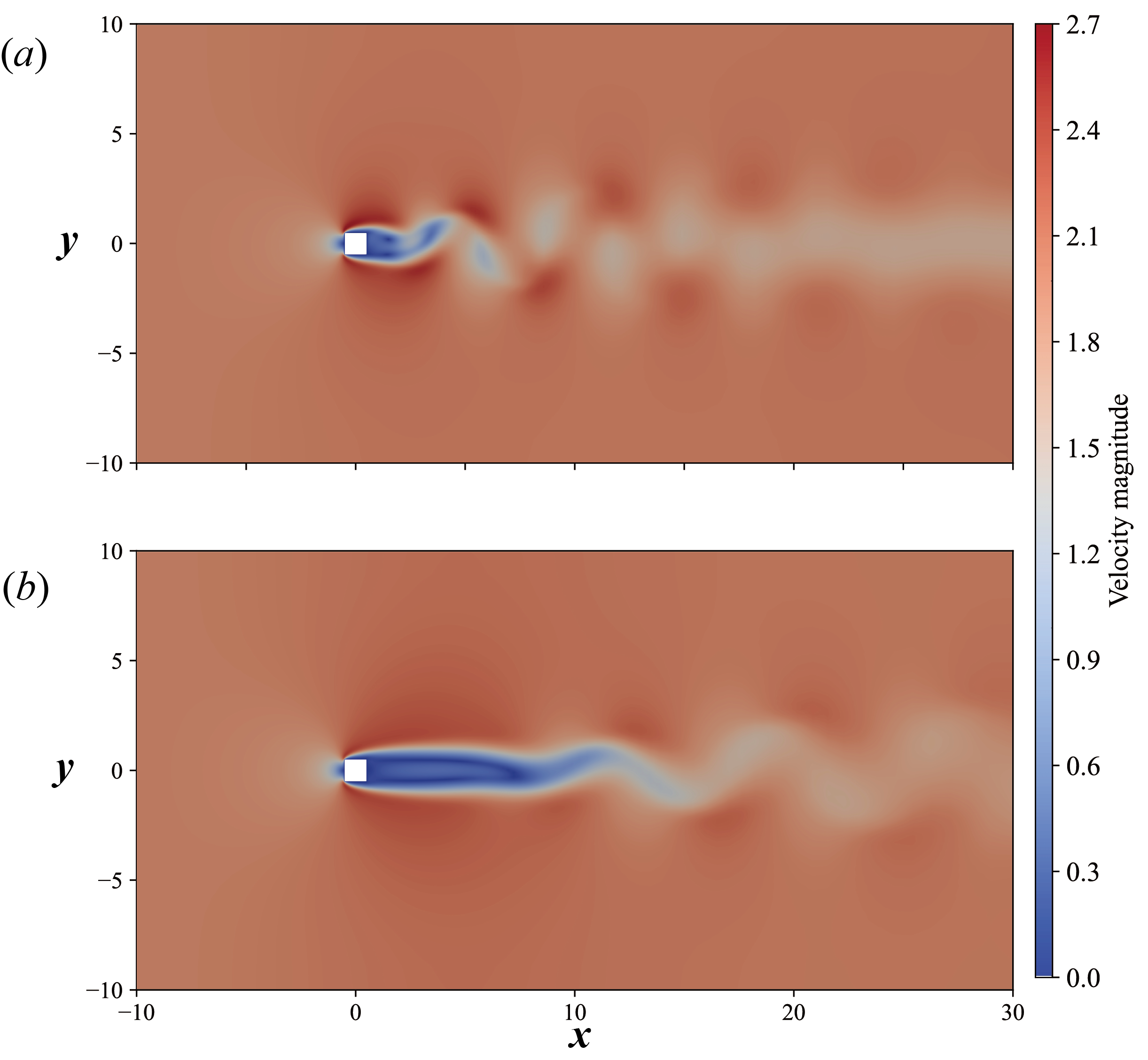} 
\caption{Comparison of representative snapshots of the velocity field around the square cylinder (a) without and (b) with active flow control.} 
\label{fig:square_comparisonU}
\end{figure}

Figure \ref{fig:square_comparisonU} shows the instantaneous flow field of the square cylinder with or without active flow control. Similarly, the effect of the controlled jet flow on the mitigation of aerodynamic force of the square cylinder could be elaborated in terms of the flow pattern. In Figure \ref{fig:square_comparisonU}(a), flow separates at the windward corners of the cylinder, and forms large-scale vortex in the near wake of the cylinder. The large-scale vortex alternatively sheds at two sides of the square cylinder, forming a clear vortex street. This alternative vortex shedding pattern directly results in the fluctuations in both drag and lift coefficients shown in Figure \ref{fig:squarecdcl}. In contrast, the vortex shedding for the square cylinder with active flow control occurs far away from the cylinder. That is to say, the effect of the alternative vortex shedding on the fluctuation of aerodynamic forces is indirect and hence very weak. That is why the fluctuation of both drag and lift coefficients is suppressed dramatically. On the other hand, the reduction in the mean drag coefficients results from the elongated wake. The elongated wake implies a reduced curvature of the shear layer and corresponds to an increased pressure at the rearward cylinder side. This increased pressure causes a smaller drag for the cylinder with active flow control.

\section{Conclusion and future work}
This study has developed a flexible, scalable, and user-friendly platform, DRLinFluids, for connecting reinforcement learning with fluid mechanics, and two case studies have been tested for validation. DRLinFluids uses one of most used open-source CFD software, OpenFOAM, as the environment to interact with the agent in different reinforcement learning algorithms from a built-in or a custom DRL backend. To share data and call different modules flexibly, DRLinFluids leverages regular expressions to handle OpenFOAM dictionary files.

The use of DRLinFluids successfully carries out a jet actuation control policy, which remarkably reduces the drag and lift force acting on the circular and square cylinders, and suppresses vortex shedding in their wake region. To find the optimal feedback control, the policy network is continuously updated during interaction to maximize the reward function, which takes both drag and lift coefficients into consideration. A significant drag reduction by up to about $8\%$ and $13.6\%$ occurs for circular cylinder and square cylinder respectively. Besides, the lift fluctuations are also suppressed to a very low level and the regular vortex shedding has been destroyed by the jet flow. It is noteworthy that similar control strategies are obtained despite two different reinforcement learning algorithms being adopted in the two cases. The jet velocity starts at a relatively large jet flow, fluctuates weakly, and eventually reaches a very low level. The corresponding ratio of the jet velocity to the incoming wind velocity is only 0.58$\%$ and 3.443$\%$ for the circular cylinder and square cylinder cases respectively. Apparently, the DRL agent has found an effective and efficient solution to maintain a small drag and a very weak lift for the cylinders. It shows that under a low Reynolds number ($Re=100$), active flow control developed based on DRLinFluids can effectively eliminate the adverse effect of vortex shedding on drag and lift of both circular and square cylinders with only a small amount of hyper-parameter tuning required.

These two validation cases demonstrate the effectiveness and reliability of DRLinFluids in DRL-based active flow control. We demonstrate the scaling of the training to both parallelization of the CFD simulation itself through the OpenFOAM framework, and the higher-level parallelization of the DRL episodes gathering by using several environments in parallel. This double scaling is critical for further applications, where increasingly complex flows are considered. More generally, since OpenFOAM has powerful application functions in CFD, we believe that DRLinFluids have a much broader application outlook in the field of fluid mechanics, which can bridge the gap between cutting-edge DRL algorithms and traditional fluid mechanics research or rapid implementation of DRL applications in the relevant industries. DRLinFluids is, therefore, an important step towards the application of DRL control to full-scale, realistic engineering configurations. Next steps in the development of DRLinFluids include the development of 3D active flow control benchmarks, testing on large scale supercomputers, and the development of automated methodologies to define invariants and symmetry-aware strategies that take advantage of the structure of the dynamic system to control, which are the key elements needed to apply DRL on increasingly complex flow control problems.

\section{Acknowledgement}
This study is supported by National Key R\&D Program of China (2021YFC3100702), National Natural Science Foundation of China (52108451, 12172109 and 12172111), Shenzhen Science and Technology Innovation Commission (GXWD20201230155427003\-20200823230021001), Shenzhen Key Laboratory Launching Project (ZDSYS20200810113601005), and Guangdong-Hong Kong-Macao Joint Laboratory for Data-Driven Fluid Mechanics and Engineering Applications (2020B1212030001)
and by Natural Science and Engineering grant GD-2022A151501149 of Guangdong province, China.

\section*{Appendix}

The DRLinFluids framework is released under an open source license on Github, at the following URL: \url{https://github.com/venturi123/DRLinFluids} [NOTE: the page is currently an empty placeholder, and the actual code will be released upon publication of the manuscript in the peer review literature]. We invite all users to further discuss and ask for help directly on Github, through the issue system, and we commit to helping develop a community around the DRLinFluids framework by providing in-depth documentation and help to new users.

\bibliographystyle{unsrtnat}
\bibliography{references}  

\end{document}